# Interdisciplinary collaboration in research networks: Empirical analysis of energy-related research in Greece

**Georgios A. Tritsaris[1,*], and Afreen Siddiqi[2,3]**

[1]*Harvard John A. Paulson School of Engineering and Applied Sciences, Harvard University, Cambridge, MA 02138, USA*

[2]*Belfer Center for Science and International Affairs, John F. Kennedy School of Government, Harvard University, Cambridge, MA 02138, USA*

[3]*Institute for Data, Systems, and Society, Massachusetts Institute of Technology, Cambridge, MA 02139, USA*

*Corresponding author. georgios@tritsaris.eu

## Abstract

Technological innovation is intimately related to knowledge creation and recombination. In this work we introduce a combined statistical and network-based approach to study collaboration in scientific authorship. We apply it to characterize recent research efforts in renewable energy technology and its intersections with the domains of nanoscience and nanotechnology with focus on materials, and electrical engineering and computer science in Greece and its broader European and international environment as a case study. Using our methods we attempt to illuminate the processes which underlie knowledge creation and diversification in these research networks: a (positive) relationship between expenditure on research and development and the extent and diversity of team-based research at the intersections of the three domains is established. Our specific findings collectively provide insights into the collaboration structure and evolution of energy-related research activity in Greece, while our methodology can be used for evidence-based design, monitoring, and evaluation of interdisciplinary research programs.

# 1. Introduction

Sustained development and deployment of clean energy technologies is expected to make a significant contribution towards climate change mitigation goals as the energy sector generates around two-thirds of global greenhouse gas emissions (International Energy Agency 2016). Fostering technological innovation in energy is especially challenging. The scale and complexity of the modern energy and technological systems necessitate national policy and state funded efforts along with international cooperation and research collaboration for devising innovative solutions to the energy challenge (UNESCO 2015).

Innovation may be interpreted as a process in which an organization or individual creates new knowledge by continuously recombining existing knowledge, and uses this knowledge to reach desired goals (Audretsch & Feldman 1996; Nonaka 1994; Weitzman 1998). Within this context, innovation (and by extension economic growth) are intimately related to knowledge creation and diversification (Breschi et al. 2003; Romer 1986, 1990). Effectively solving complex scientific and technological problems often involves research at the intersection of different science and technology (S&T) domains. An interdisciplinary approach is often pursued to facilitate the integration of concepts and information from differentiated, albeit related, knowledge sectors (Cummings & Kiesler 2005; Jacobs 2009). As a matter of fact, many recent innovations in energy devices and systems have been achieved with research at the intersection of the domain of energy technology and the domains of nanoscience and nanotechnology, and electrical engineering and computer science. Cases in point are the search for novel nanostructured materials for cost-effective solar energy harvesting, the use of computer simulations to accelerate the design of more efficient wind turbines, the simulation and optimization of hybrid renewable energy systems, and the development of intelligent support systems to assist the regulation of distribution networks (Arico et al. 2005; Baños et al. 2011; Bernal-Agustín & Dufo-López 2009; Bottasso et al. 2014; Grätzel 2005; Huynh 2002; Wang 2004; Yu et al. 2015). The overlap between these three S&T domains has also been identified in previous studies that have tracked research on renewable energy and nanotechnology (Arora et al. 2013; Kajikawa et al. 2008).

Knowledge creation can be induced by information brought into an organization or by synthesis of information by such boundary-spanning actors (Aldrich & Herker 1977). Research systems when become isolated from external influences risk moving to a state of internal compromise between units and subsequent stagnation. By maintaining links with the environment (or multiple environments), active intermediaries in research, development and innovation (RDI) can support the sustainability of the system, create opportunities for entrepreneurship, and prevent structural lock-in. In order to devise policy instruments to promote energy technology innovation, it is important to know the type and extent of related research so that the allocation of RDI funds can be strategically directed. For instance, cost-effectiveness can be improved by informed leveraging of the S&T knowledge created and distributed within and across various sectoral, regional and national research networks. Mapping international research collaboration is of particular interest owing to the researchers' role in external representation of national research and in absorbing and recombining information on the national-international boundary. Advances in methods that aim to define the extent and intensity of boundary-spanning research and illuminate the processes which underlie knowledge creation, recombination and diversification are crucial for the design of effective technology innovation policies. The main contribution of this work is a novel approach to studying collaboration and interdisciplinarity in scientific authorship that draws from the fields of infor-mation retrieval, statistics and network analysis. Statistical analysis of information about scholarly publication of research has served as a transparent means to improve decision-making in knowledge management (Cronin & Sugimoto 2014). Defining the scope of a S&T domain based on scholarly publication is a methodological challenge. We embed the collection of publications with simple search terms in a workflow for information retrieval that serves to enlarge an initial set of core publications by a form of query expansion (Bettencourt & Kaur 2011; Mogoutov & Kahane 2007). On the basis of the collected publications, we construct and analyze coauthorship networks to map boundary-spanning research, highlighting research clusters (teams) as the elementary unit as opposed to studies of team formation and evolution which examine the internal composition of teams and their size (Guimerà et al. 2005; Milojević 2014). In doing so, we complement previous work that aims to define diversity-based indicators of interdisciplinarity (Rafols & Meyer 2010).

Our approach offers a multi-level view of knowledge organization on the basis of the research interests and collaboration partners of individual scientists (M. E. J. Newman 2004; de Solla Price 1965). We use it to map research efforts in renewable energy technology (RET) and its intersections with the domains of nanoscience and nanotechnology with focus on materials (NNM), and electrical engineering and computer science (EECS). We apply this approach to Greece and its broader European and international environment as a case study. Greece is considered a relatively moderate innovator, with research and development (R&D) intensity (0.96% in 2015) below the EU-28 average, but an especially strong performer in international scientific collaboration. There remains a substantial divide between the best and worst performing national innovation systems (Edquist 1997; Sharif 2006) in EU according to the European Commission's Innovation Union Scoreboard, a measurement framework developed to assess convergence among member states and improvement in overall European innovation performance (Arundel & Hollanders 2008; Hollanders & Es-Sadki 2016).

With focus on Greece, we examine the content, organizational make-up and geographical trace of scientific collaboration and how these have evolved over the sixteen-year period 2000-2015. Although the peculiarities of any particular S&T domain are well-known to domain experts, here we concentrate attention on trends in publishing output at the national level and beyond. Rather than offering policy prescriptions we use the case of energy-related research in Greece to provide working examples of how the proposed approach can be used to identify points of intervention for RDI policy. An important question we attempt to answer using our methods is to which extent independent energy-related research efforts have contributed to shape existing knowledge diversity into a distinct field of inquiry. Our specific findings collectively provide insights into the collaboration structure and evolution of energy-related research activity in Greece and contribute towards an improved understanding of the Greek innovation system. Our approach can also be used for other regions, and more generally for operationalizing boundary-spanning research to design, monitor, and evaluate interdisciplinary research programs based on empirical evidence and inform RDI policy for energy technology or other S&T domains.

The discussion is organized as follows. In Section 2 we introduce general methods for data collection and analysis of coauthorship networks, including boundary-spanning. In Section 3, each of the three S&T domains is defined by characterization of scholarly publication of research. In doing so, we examine international collaboration and its effect on local (regional) coordination. In Section 4, single-domain and boundary-spanning coauthorship networks are constructed and analyzed using the methods introduced in Section 2. For instance, using binary operations on the coauthorship networks we search for authors active at the intersections of the three S&T domains, and we establish a (positive) relationship between number of research teams and expenditure on R&D. Finally, the broader implications of our specific findings are discussed in Section 5.

## 2. Data and methods

Bibliometric studies, or bibliometrics, is an approach to constructing indicators of RDI output which is based on the statistical analysis of information about scholarly publication of research such as articles in scientific journals and patents supporting creation and flows of knowledge. Owing to the easy availability of publication and citation data, bibliometrics has served within the appropriate contextual framework as a transparent and cost-effective means to improving the quality of expert decision-making and planning in knowledge management (Cronin & Sugimoto 2014). This study covers the sixteen-year period 2000-2015 and relies on publication and citation data available through the *Web of Science* (*WoS*) platform and its core selection of citation indexes and databases. Hereafter, we will refer to the complete set of retrieved records as the "collection" ("EL" label in figures).

### 2.1 Information retrieval

Publications within the RET and NNM domains were identified in the collection using keyword-based queries with equal weight on publication titles, abstracts and associated keywords (Manning et al. 2008). There is no universal approach to delineating a S&T domain and obtaining a set of publications through searches in citation databases within a particular domain is a methodological challenge (Arora et al. 2013; Bettencourt & Kaur 2011; Manzano-Agugliaro et al. 2013; Mogoutov & Kahane 2007). For instance, in Betten-court & Kaur (Bettencourt & Kaur 2011) a collection of publications in sustainability science was assembled on the basis of simple search terms while the methodology defined in Mogoutov & Kahane (Mogoutov & Kahane 2007) is an example of an evolutionary approach used to augment a core set of keywords for nanoscience and nanotechnology. We adopted the general principles of a) internal consistency, in order to enlarge an initial set of core publications by a form of query expansion that aims to reduce expert intervention, and b) reproducibility, on the basis of a clearly defined workflow for information retrieval. The details of the workflow are provided in the supporting information. Briefly, for each of the two S&T domains a set of handpicked terms based on expert judgement was used to retrieve an initial set of core publications and the keywords associated with them. Separate sets of publications were retrieved using these keywords, one at a time. A keyword was added to the initial set of handpicked terms if at least one of the latter appeared among the most frequent keywords associated with the working set of publications.

### 2.2 Network modeling and analysis

Networks offer a concise means of representation of connections or interactions between the parts of a technological, informational, social or other system which can be studied by formal graph theory and network analysis (M. Newman 2010). We used the collection of publications for each S&T domain to build coauthorship networks in which the vertices describe authors (or researchers) and two authors are connected by a link if they have coauthored at least a certain number of scientific publications. Mathematically the networks are represented as undirected graphs. For network analysis we used *NetworkX* (version 1.11), a Python open-source library for the creation and study of complex networks (Hagberg et al. 2008).

The structure and temporal evolution of various coauthorship networks were characterized using common network (graph) methods. The density $d_G$ for a network $G$ with $n$ vertices and $m$ edges was calculated as

$$d_G = \frac{2m}{n(n-1)}.$$

The density of a graph without edges is 0, and 1 for a complete graph. We also used betweenness centrality, $c(v)$, as a measure of centrality based on shortest paths for some vertex $v$ in a given network:

$$c(v) = \sum_{i,j \in V} \frac{\sigma(i,j|v)}{\sigma(i,j)},$$

where $V$ is the set of vertices, $\sigma(i,j)$ is the number of shortest $(i,j)$-paths, and $\sigma(i,j|v)$ is the number of those paths passing through $v$ other than $i, j$. In the context of our investigation, betweenness centrality offers a measure of 'importance' that relates to where a researcher is located with respect to knowledge flows in the coauthorship network, that is, how frequently the researcher is found on the shortest path between a pair of other researchers. Higher (lower) values for betweenness centrality suggest more (less) significant role in mediating knowledge flows in the network.

In order to identify authors who create bridges between any two or more S&T domains, we performed binary operations on the corresponding coauthorship networks. The intersection, $G_{ij}$, of two networks $G_i = (V_i, E_i)$ and $G_j = (V_j, E_j)$ is given as

$$G_{ij} = (V_i \cap V_j, E_i \cap E_j),$$

where $V_{i,j}$ are vertices and $E_{i,j}$ are edges in the networks.

For efforts in R&D to be self-sustaining and to maintain or develop scale and scope, they need to reach and maintain critical mass while they remain diversified enough to foster innovation. We introduced two different time-dependent indicators to quantify knowledge diversity in any combined network describ-ing all possible intersections of three or more domains. Both indicators are based on a simple counting of researchers (vertices) at these intersections. A variance-based diversity indicator is defined as

$$V_t = 1 - \sum_k p_{k,t}^2,$$

and an entropy-based diversity indicator is defined as

$$H_t = \sum_k p_{k,t} \log_{10} \frac{1}{p_{k,t}},$$

where $p_{k,t}$ is the fraction of authors who have published at an intersection $k$ of two or more S&T domains over a given period $t$. Higher values for the indicators suggest higher diversity in the knowledge stock and more opportunities for assimilation and production of novel knowledge.

The workflows developed for publication and citation management for the purposes of this study support queries such as *"retrieve all publications to which at least one author from network G has contributed"*. Such a query would enable detailed characterization of scholarly work of selected authors and their collaborators within any one of the S&T domains or their intersections.

# 3. Scientific and technological domains

An interdisciplinary approach (or multidisciplinary; here we use the two terms interchangeably) to energy research has often been pursued in order to create opportunities for the synthesis of concepts and information from different knowledge sectors and create new knowledge for solving complex problems in energy science and technology. In this work we concentrate attention on the domain of renewable energy technology and its intersections with the domains of nanosci-ence and nanotechnology with focus on materials, and electrical engineering and computer science.

## 3.1 Definitions

We describe how each S&T domain was defined for the purposes of our analysis in the following.

**RET.** Scientific publications within the RET domain were identified in the collection using keyword-based queries as described in Section 2. The query terms used describe renewable energy sources, related technologies for energy conversion, and fuels. Our description of the RET domain was elected to be at the device or system level in accordance with previous evaluations of pertinent literature (Kajikawa et al. 2008; Manzano-Agugliaro et al. 2013) and in order to match the content and typical level of abstraction of priority sectors in EU RDI policy and funding schemes: improving energy efficiency and supporting energy-related research and technological development are closely related dimensions of the current EU policy framework for climate and energy, alongside the promotion of cooperation among its members (European Commission 2014), while clean energy is among the cross-cutting issues mainstreamed in the EU's current funding program for RDI, Horizon 2020, but also in similar funding schemes in the European neighborhood.

**NNM.** The NNM domain exhibits a high degree of interdisciplinarity, encompassing aspects of physics, chemistry, biology, materials science and engineering, and other theoretical and applied academic disciplines (Huang et al. 2011; Porter & Youtie 2009). Moreover, nanotechnology and advanced materials are considered general-purpose technologies which enable innovations in other domains (Bresnahan & Trajtenberg 1995; Jovanovic & Rousseau 2005). The query terms used to define the NNM domain describe common objects of study and their morphology and closely match compiled lists used in previous evaluations of the nanoscience and nanotechnology literature (Arora et al. 2013; Mogoutov & Kahane 2007). The complete list of query terms for the RET and NNM domains is provided as supporting information.

**EECS.** Like NNM, electrical engineering and computer science (EECS) are enabling other sectors as in the case of e-infrastructures for scientific computing, shared social data resources, or the activity of public sector organizations. Being an established academic discipline, a large number of universities worldwide have traditionally offered degrees in EECS. To capture the knowledge base of this particularly broad sector for the purposes of our analysis, instead of using a keyword-based search as previously, publications were simply retrieved on the basis of subject areas assigned to them in *WoS*, namely "Electrical and Electronic Engineering" and "Computer Science". Considering electrical engineering and computer science together bridges the gap between hardware and software systems.

In addition to receiving sustained support through European thematic research funding programs, the elected three S&T domains have been identified among the priority sectors within the context of research and innovation strategy for smart specialization in Greece. In no case we attempt to distinguish between basic and applied research (Godin 2006).

## 3.2 Publishing output

Although our analysis approach is independent of the bibliometric set used, the collection will be inevitably dependent on the elected protocol for record retrieval. Before studying the interface between the S&T domains therefore it is necessary to offer a detailed characterization of each domain and this way define their scope beyond associated keywords. Regarding the performance of the Greek innovation system, emphasis has traditionally been placed on measures of innovation inputs such as R&D expenditure and less on indicators describing output which can support evidence-based monitoring of research activity towards stated goals.

## Publishing volume

The yearly number of publications in the collection increased between 2000 and 2009 by around 500 publications each year and after a brief decrease during the years 2010 and 2011 it has risen albeit more slowly to reach 11664 in 2015 (Figure 1a). Nevertheless the number of cited publications has been increasing. Here, we consider a publication as cited if it has received at least one citation in the year of publication or the following year regardless of total number of citations. The yearly number of author contributions has been increasing during the entire time period, with 3 or 4 contributions per publication on average over the period 2010-2015. The corresponding number of contributions at the level of individual organizations has been 1 or 2. A similar trend is observed for the three S&T domains although the decrease in publication production during the years 2010 and 2011 is much more pronounced for the EECS domain (Figure 1a, "EECS").

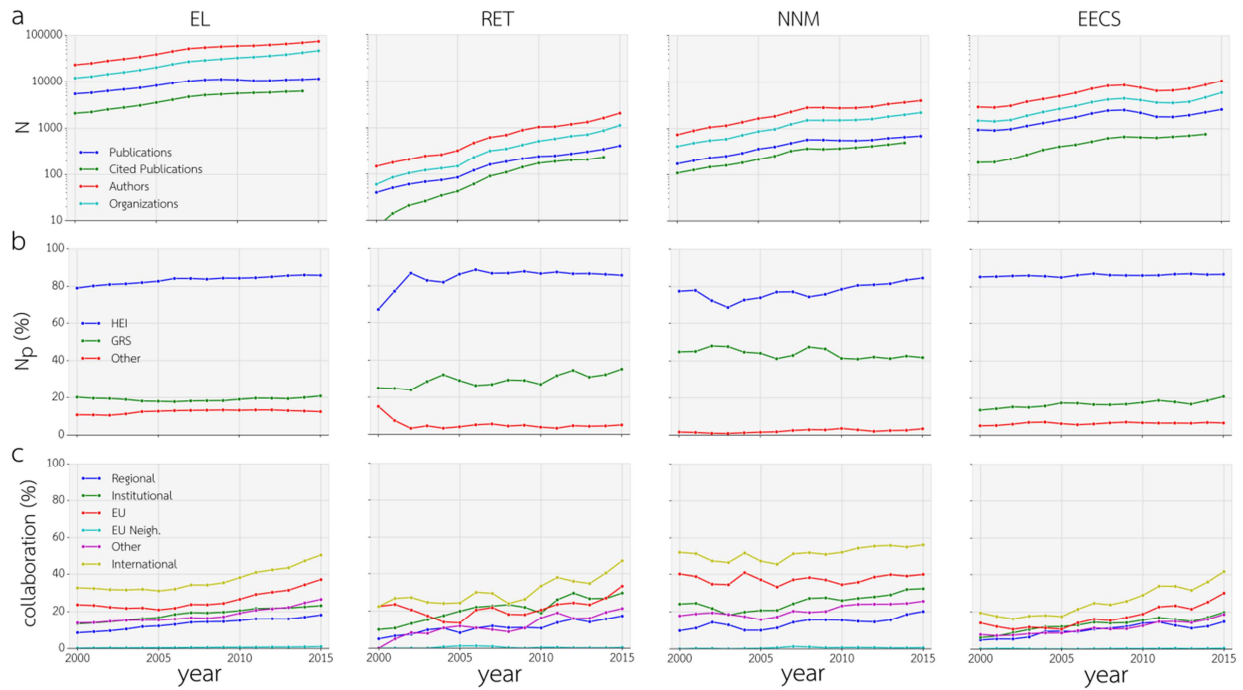

Figure 1. a) Yearly number of scientific publications (see text for definition of "cited"), author contributions, and organization contributions. b) Share of scientific publications by type of organization. Contribution of higher education institutions (HEIs) is shown in blue, and contribution of governmental research centers (GRCs) is shown in green. c) Share of publications by type of collaboration in authorship.

## Disciplinary composition

We examine next the disciplinary composition of each S&T domain in more detail. Figure 2 shows the absolute share (whole counting, shown in labels; sum over 100%) and relative share (slice size by whole-normalized counting; sum of 100%) of publications by

OECD's taxonomy for S&T fields (Frascati manual classification scheme) on the basis of *WoS* subject categories. The two measures do not coincide because each publication can be assigned to multiple subject categories in *WoS* resulting to some overlap between Frascati fields in over 40% of the collection. Natural Sciences, and Engineering and Technology are the two fields which are generally represented more strongly. Over the period 2010-2015, around half (53.2%; whole counting) of national publishing output has been in the field of Natural Sciences, one-third in Engineering and Technology (32.8%) and one-third in the Medical and Health Sciences (31.9%). Within the NNM and EECS, between 70% and 82% of publications are in Natural Sciences while within the RET and EECS domains the field of Engineering and Technology was assigned to 70%-78% of publications. Across the three S&T domains the Medical and Health sciences are underrepresented.

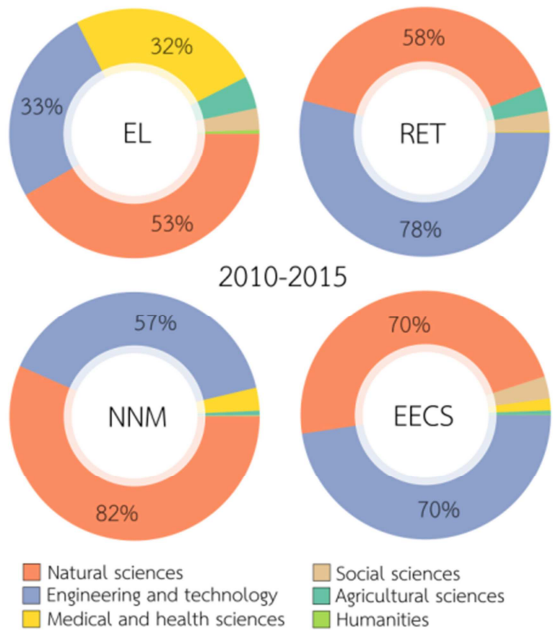

Figure 2. Absolute share (whole counting, shown in labels) and relative share (slice size by whole-normalized counting) of scientific publications over the period 2010-2015 by OECD's taxonomy for science and technology fields (Frascati manual classification scheme).

Overlaps between the knowledge bases of the three domains are apparent at the coarse level of classification by *WoS* subject categories: The three most assigned *WoS* subject categories over the period 2010-2015 are Energy and Fuels, Physical Chemistry, and Multidisciplinary Materials Science within the RET domain, and Multidisciplinary Materials Science, Applied Physics, and Physical Chemistry within the NNM domain. Decomposition of the EECS domain into *WoS* subject categories reveals emphasis on Theory and Methods, and Information Science.

### Organizational composition

We shift attention to the organizational composition of research activity. Only national higher education institutes (HEIs) and governmental research centers and laboratories (GRCs) are considered in the following (see supporting information for more on the classification scheme for organizations). Across the entire collection (period 2000-2015), the contribution of HEIs to scientific publishing output has been between 79% and 86% (whole counting) following an upward trend, while the contribution of GRCs has been between 18% and 21% (Figure 1b). Within the GRC sector, research centers supervised by the General Secretariat of Research and Technology (GSRT) have had a share between 76% and 79% of organization contributions. The EECS domain follows a similar trend, while the contribution of GRCs has been twice as high within the RET and NNM domains.

### 3.3 Scientific collaboration

Next we search for trends in scientific collaboration. A key assumption of EU's present RDI policy is that there exist increasing returns to scale arising from knowledge spillovers through collaboration in R&D that lay the foundation for further knowledge creation, diffusion and use. On the basis of the authors' institutional affiliations we distinguish the following (possibly co-occurring) types of collaboration: between a) at least two Greek organizations which are classified into different types ("Institutional" in Figure 1c), b) at least two different administrative divisions (or regions) of Greece according to NUTS2 codes and names ("Regional"), c) Greece and at least one other EU-28 member state ("EU"), d) Greece and at least one neighboring country in Western Balkans ("EU Neigh."),[i] and between e) Greece and at least one country which is not described by (c) or (d) ("Other"). We refer to (c), (d) and (e) together as international collaboration ("International"). By inspection of Figure 1c, which shows the share of publications by type of collaboration, we identify a trend of increasing collaboration in authorship, within and beyond national borders, and across all three S&T domains.

## Interorganizational collaboration

Interorganizational collaboration within Greece, as defined above, has increased across the entire collection with 22% of all publications between 2010 and 2015 to be a result of such collaboration. The same trend is observed within all three S&T domains, with RET showing the largest increase, from 10% in 2000 to 30% in 2015 (Figure 1c, "RET"). HEIs have contributed to more than 96% of all interorganizational collaborations within all three S&T domains (whole counting), which reflects HEIs' large share of publishing output in general, while GRCs' contribution within the RET domain has been around 80% (79% over the period 2010-2015), around 90% (85%) within the NNM domain, and around 60% (59%) within the EECS domain.

## Regional collaboration

Collaboration between different types of organizations may entail collaboration between different regions. Regional specialization and diversification is considered a crucial driving force in further enhancing the potential for innovation because close geographic (physical) proximity and differentiated, albeit related, knowledge bases are expected to facilitate knowledge creation and technological learning (Audretsch &

Feldman 1996; Cooke 2001; Frenken et al. 2007; Hidalgo & Hausmann 2009). We found that regional collaboration was twice as common in interorganizational collaborations across the entire collection (44% for 2010-2015 publications; compare with Figure 1c). Over the period 2010-2015, the three regions that have participated in most interregional collaborative research across the entire selection are Attica with a share of 35% of contributions at the level of institutional affiliations (whole-normalized counting), Central Macedonia with 17%, and Western Greece with 10%. However, these calculations do not take into consideration the relative research capacity of each region. After controlling for differences in research inputs different regions emerge as research hubs within each of the three S&T domains, as shown in Figure 3. The coloring scheme codifies the relative number of contributions per gross domestic expenditure on R&D (GERD) of each region within each domain in the year 2013 (see supporting information for data; we could not readily identify data for other years): shades of red or blue indicate higher or lower contribution than the national average for the same year, and grey indicates insignificant output (less than ten contributions in total). In interregional copublication, Western Greece and Western Macedonia are found to be significant contributors within the RET domain, while Western Greece and Epirus are the major contributors within the NNM domain. Regional collaboration within the EECS domain is found to be the most diverse, which reflects the broad scope of our operational definition, with Central Greece (NUTS1) showing above average copublishing output. Geographic proximity generally favors regional collaboration since scientific and technological expertise can be "sticky", that is, hard to acquire, transfer and use in new locations, but it is not the sole determinant of regional clustering as the case of Crete within the NNM domain suggests. The trends do not change when FTE is used as a descriptor of research inputs instead of GERD. Moreover, the level of publishing output relatively to the national average may indicate propensity towards either specialization or variety and not necessarily over- or underperformance. Such insights should be useful for informing the design of RDI policies which aim to promote linkages between regions of complementary research orientations. We will return to this point later.

[i] Albania, FYR Macedonia, Kosovo, Montenegro, Serbia, Bosnia and Herzegovina

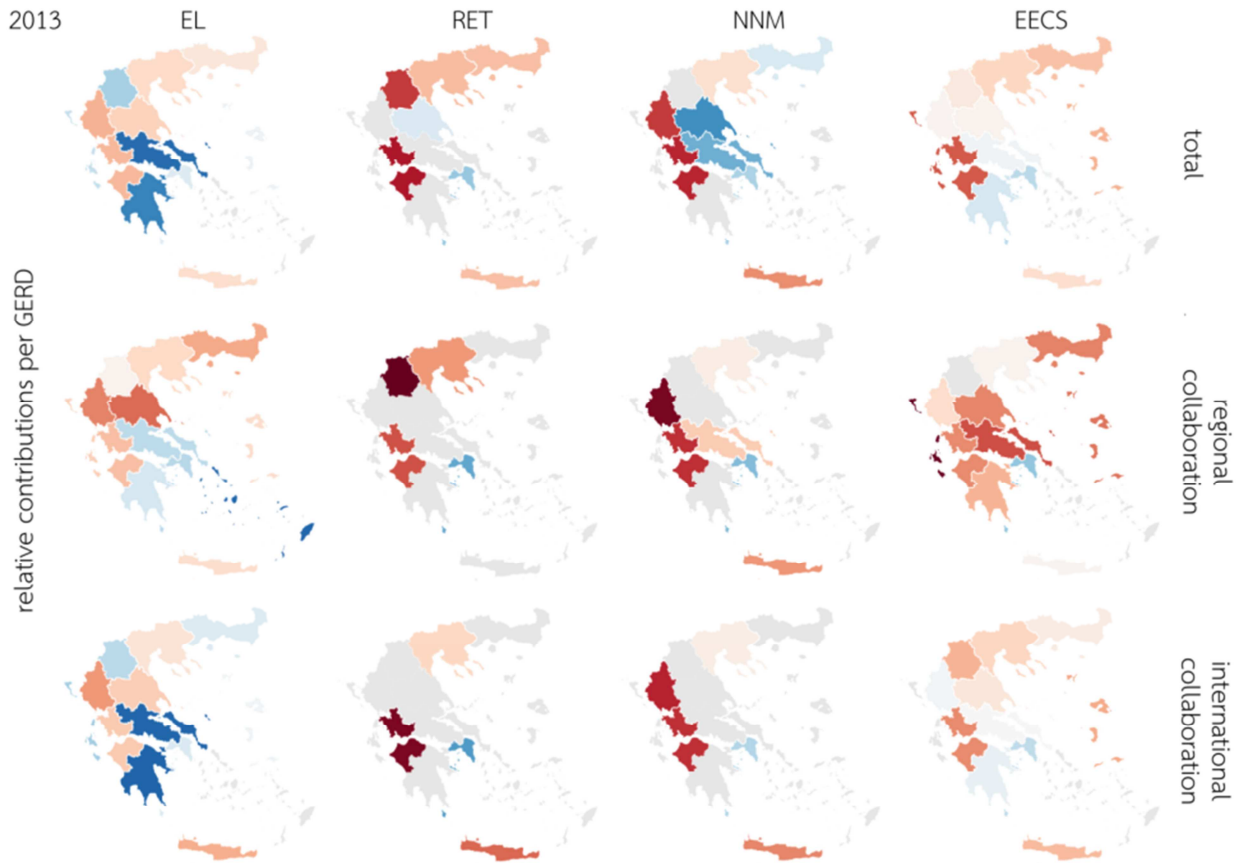

Figure 3. Total contributions of each region in authorship relatively to the national average (top row), contributions in interregional collaborations (middle row), and contributions in international collaboration for the year 2013. Red (blue) shades signify more (less) contributions than the national average.

## International collaboration

The growing complexity of modern technological problems has also introduced an international dimension in RDI policy that encourages exchange of knowledge via international linkages and collaborations (Luukkonen et al. 1992; UNESCO 2015). Collaboration beyond national borders has accelerated after 2010, with European collaboration more common than collaboration between organizations or regions within Greece (Fig-ure 1c). The NNM domain stands out as especially outward-looking. With respect to contributions at the level of institutional affiliations, Attica has had the highest participation in international research, 30%-43%, across the three S&T domains over the period 2010-2015. Attica being the most represented in regional and international collaboration reflects its status as a socio-economic center. In terms of contributions per GERD (2013 levels), major contributors in international collaboration within the RET and NNM domains were also important contributors in regional collaboration (Figure 3, "RET" and "NNM"). In this case therefore increased international collaboration is associated with more extended local coordination, a fact which can have implications for the design of multi-target policy instruments. The relative contribution between the core and the periphery within the EECS domain is reversed with respect to regional collaboration (Figure 3, "EECS"). Moreover, although Crete was visible worldwide in all three S&T domains, it did not stand as a strong collaborator between regions within the RET and EECS domains. This could be attributed to the geographic isolation of the region. Across the entire collection, the leading countries in terms of total number of coauthored publications are USA (11% share of contributing countries for 2010-2015), UK (10%), Germany (8%), Italy (7%) and France (6%), which have been traditionally strong in science. The share of each of these countries has been declining the entire time period while collaboration partners have become more diverse. Regarding collaboration between Greece and its European neighbors outside EU-28, small number of scientific copublications do not allow for a reliable and detailed statistical analysis (Figure 1c). If public policy intervention is desirable, incentives for rewarding collaboration however could be created by means, for example, of thematic research funding. Moreover, researchers affiliated with at least one Greek and one international organization can facilitate the transmittance of knowledge across national borders: around 20% of all authors in scientific publications across the three S&T domains are identified to work at this boundary (Figure 1c).

Additional findings with respect to national publishing output and the research impact of the S&T domains is provided as supporting information.

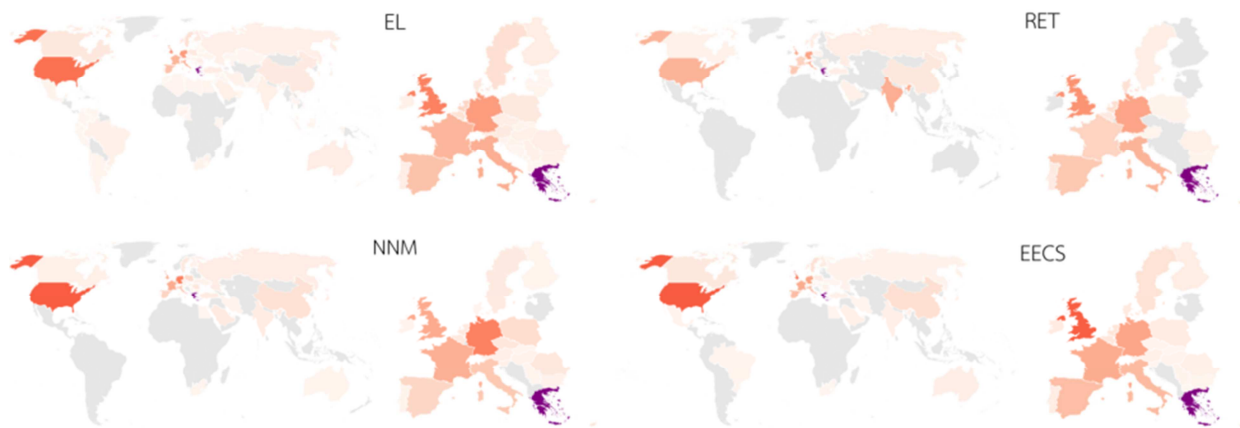

Figure 4. Countries contributing in scientific publication over the period 2010-2015. Darker (lighter) shades of red signify higher (lower) share of publications. The map of EU-28 and neighboring countries is shown in greater detail.

## 4. International collaboration networks

Research systems are open systems and knowledge can be transferred to or absorbed from RDI actors outside. Social proximity in a scientific or technological community or organization can in part explain such knowledge externalities (Aguiléra et al. 2012; Boschma 2005), while a scientific community can be understood as a collective of individuals producing knowledge in a particular scientific field and who interact more closely with each other than with other unrelated individuals even in close physical proximity (Breschi et al. 2003).

In this section we examine scientific collaboration beyond national borders by constructing and analyzing coauthorship networks.

## 4.1 Single-domain research

Having characterized in detail each of the three S&T domains in the previous section, we proceed to map scientific collaboration using coauthorship networks in which the vertices describe authors (or researchers) and the links correspond to coauthorship weighted by the number of collaborations in a given time period. Only these publications which are the product of an international collaboration and which have received at least one citation the year of publication or the following year were considered. We describe networks of two or more copublications between linked authors. The structure and temporal evolution of the coauthorship network for each S&T domain were quantified using common network (graph) descriptors, which are presented in Table 1. Increased international contribution in authorship is identified in the collaboration networks for all three S&T domains over the period 2008-2013 in comparison to the period 2005-2010, with more authors and linkages (Table 1). The average number of collaborators has been higher for the NNM domain on average, while the collaboration network for the RET domain has been denser. All three collaboration networks have become sparser as their size increased.

| | RET | | | NNM | | | EECS | | |
|---|---|---|---|---|---|---|---|---|---|
| | 2002-2007 | 2005-2010 | 2008-2013 | 2002-2007 | 2005-2010 | 2008-2013 | 2002-2007 | 2005-2010 | 2008-2013 |
| **Authors** | 102 | 197 | 323 | 669 | 952 | 1305 | 453 | 778 | 1082 |
| **Linkages** | 305 | 413 | 656 | 1664 | 2400 | 3607 | 954 | 1443 | 2289 |
| **Collaborators per author, average** | 6.0 | 4.2 | 4.1 | 5.0 | 5.0 | 5.5 | 4.2 | 3.7 | 4.2 |
| **Density** | 0.059 | 0.021 | 0.013 | 0.007 | 0.005 | 0.004 | 0.009 | 0.005 | 0.004 |

Table 1. Time-evolution of coauthorship networks for the RET, NNM and EECS domains (two or more copublications between linked authors).

## 4.2 Boundary-spanning research

Thematic research priorities which are adopted to improve cost-effectiveness may stifle creativity and innovation and undermine future developments, and so an interdisciplinary approach to research is often pursued to create opportunities in integrating concepts and information at the intersection of knowledge sectors. It is therefore desirable to be able to monitor the extent of knowledge diversification but also examine evolution from transgression of disciplinary boundaries to integration of research efforts into a distinct knowledge domain.

In the following we examine in more detail international collaboration at the intersections of the previously defined and characterized RET, NNM and EECS domains by identifying boundary-spanning actors who create bridges between the domains. Research at the intersection of the NNM and EECS domains could support energy-related research, albeit less directly. The uncertainty inherent in the innovation process is likely to increase flexibility in the specificity of boundary-spanning activities but a network-based description of research activity offers a field-level view of the knowledge profile (research interests and collaboration partners) of individual scientists. Figure 5 shows snapshot views of the resulting network, which we hereafter refer to as the boundary-spanning network and which serves as a proxy to the coordination of knowledge at the intersection of S&T domains on the national-international boundary. The vertices are color-coded with respect to the S&T domains within which a researcher has published, the vertex size is proportional to the betweenness centrality of an author in the connected graph component (or cluster) they belong to (Freeman 1977), and the edge thickness is proportional to the number of copublications between the two linked authors. The list of authors was manually inspected to ensure that each vertex corresponds to a unique author.

To minimize the influence of brief linkages, formed for example because of short-term funding opportunities, we studied the time evolution of the boundary-spanning network using sliding six-year periods first. By inspection of Figure 5 we find that among the boundary-spanning research activities, most have been focused on the intersection of the RET and NNM domains (59% of vertices over the period 2008-2013), as well as between then NNM and EECS domains (35%). Examples of boundary-spanning research include studies of nanostructured materials for dye-sensitized solar cells and simulation-aided characterization of nanocomposites, respectively. We find only very few authors to be active at the intersection of all three S&T domains, with a multiscale study of performance issues in microelectronics being one example of research at this boundary. Strong ties in the boundary-spanning network are expected to have greater value in mediating the flow of complex and dynamic information but weak ties can still provide novel information. In any case, strong ties are not a prerequisite for high centrality: a small number of researchers exist who have high centrality but are linked to each other via weak ties (i.e., sharing few copublications) or even have only a small number of connections, serving as a "quiet middle".

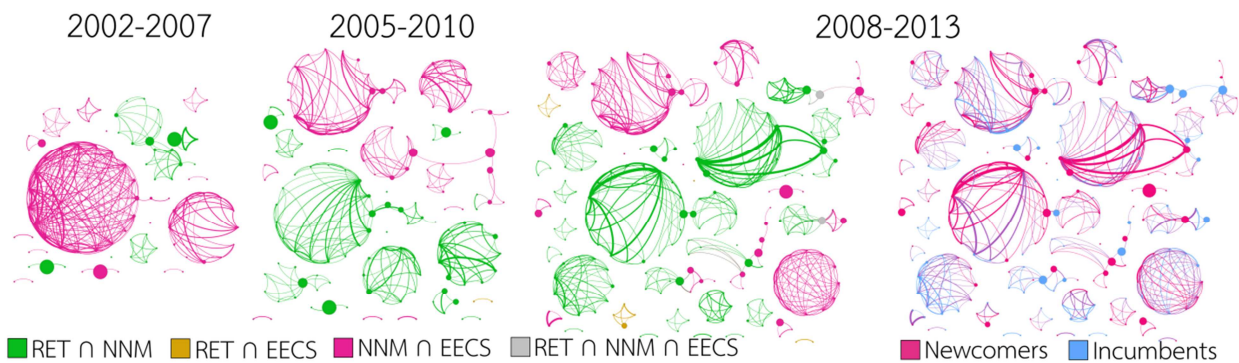

Figure 5. Structure and evolution of boundary-spanning network in scientific publication at the intersection of any two or all three science and technology domains (RET, NNM and EECS). Vertices represent researchers which are linked if they have coauthored at least two publications. The size of vertices and edges is proportional to the betweenness centrality and number of coauthored publications respectively.

In studying knowledge flows and spillovers, and other systemic aspects of innovation, clusters in the network indicate the boundaries of scientific communities and knowledge flows which define a S&T domain, regardless of geographic distance between researchers. Knowledge of such boundaries is useful in identifying the level of aggregation at which the search for correlations between research inputs, activity and outputs can be meaningful in policy design. Clustering within the boundary-spanning network is immediately obvious and puts forward the research team as the basic unit of research organization (Guimerà et al. 2005; Milojević 2014). For instance a research cluster with focus on transition metal-based nanomaterials for energy transformations at the intersection of RET and NNM was active over the period 2008-2013. These research clusters remain largely disconnected over the sixteen-year period. We observe that the boundary-spanning network has grown primarily as new authors attach to authors within existing research clusters (teams), giving an impression of how established knowledge diffuses from incumbent researchers to newcomers. Some newcomers will act as knowledge brokers who bridge previously distinct clusters or as members of new clusters. Descriptors for the network for three different time periods are provided in Table 2.

Guimerà et al. (Guimerà et al. 2005) described the role of newcomers in the self-assembly of research teams, while in Milojević et al. (Milojević 2014) the evolution of the size of scientific teams was proposed to be a multimodal process differentiating between core and extended teams. On the other hand, generative models of (collaboration) networks typically describe connected graphs (M. E. J. Newman 2004). Here, we go beyond the internal composition of teams and their size and we focus on the evolution of the broader network assuming research clusters as its elementary unit. The network has grown in number of authors and collaborations, and number and size of clusters. Guided by the continuously growing number of these clusters, we postulate that the main process which underlies the structural evolution of the boundary-spanning network is in fact gross, as opposed to thematic, expenditure on R&D. Figure 6a shows the number of research clusters, regardless of size, formed within a given time period. The time span increases from one year for the year 2000 to fourteen years corresponding to the period 2000-2013. Expenditure is described by the cumulative sum of GERD for each time period. A clear relationship exists between the two, which makes research funding a predictor of network growth, albeit not necessarily the cause. An important result is that the network comprises mostly disconnected clusters at all times –thus there will be seemingly persistent barriers in pursuing the advancement of interdisciplinarity (Jacobs 2009) in absence of suitable incentives. For example, tailored funding and research policies that create opportunities for interdisciplinary research at the intersection of all three S&T domains can create bridges between such distinct but complementary research themes towards tighter integration of research efforts.

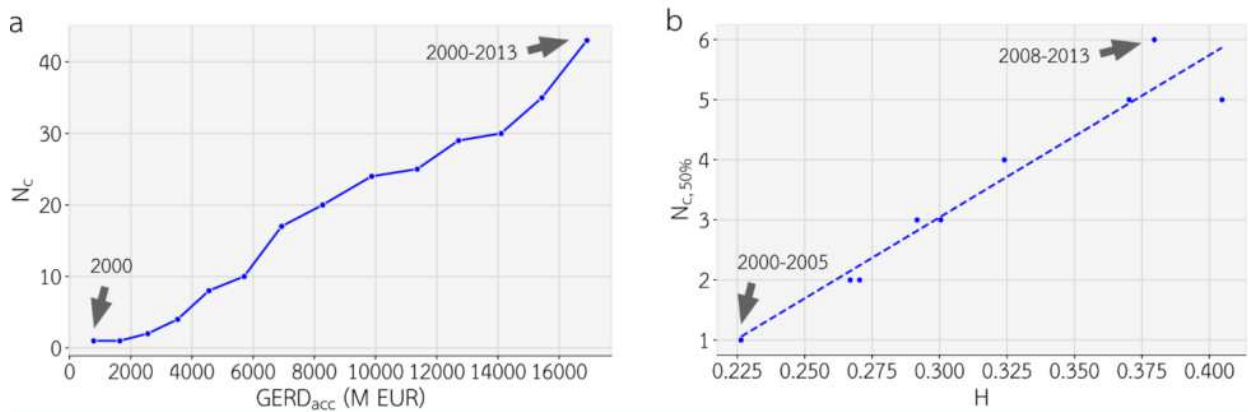

Figure 6. a) Correlation between the number of clusters (research teams), $N_c$, in the boundary-spanning network and cumulative GERD (increasing time span). b) Relationship between the number of clusters (teams) constituting 50% of the boundary-spanning network, $N_{c,50\%}$, and diversity in the network (sliding six-year periods).

For efforts in R&D to be self-sustaining and to maintain or develop scale and scope, they need to reach and maintain critical mass while they remain diversified enough to foster innovation. We use the time-dependent indicators $V$ and $H$ to quantify diversity in the boundary-spanning network. 8b shows how diversification is correlated with the number of clusters and provides a description of the relationship between diversification and integration in the knowledge network. Similarly to conceptual frameworks such as the one introduced by Rafols and Meyer (Rafols & Meyer 2010) which are also using diversity and other indicators to describe knowledge integration on the basis of low-level analytical units, we focus on independent research clusters. We find that the number of research clusters, $N_{c,50\%}$, that together comprise 50% of the boundary-spanning network on the basis of sliding six-year period has increased with the corresponding diversity indicator $H$. A fitted linear regression model quantifies the positive relationship as

$$H = (N_{c,50\%} + 5.1) \, / \, 27.0$$

Naively it could be argued on the basis of this finding that increased R&D spending would result in higher levels of boundary-spanning research. However, although this might indeed be the case the above relationship describes isolated strands of research during the entire time period –thus a distinction needs to be made between fortuitous and deliberate boundary-spanning research. A policy prescription that aims to create more opportunities for information and knowledge synthesis at the intersection of all three S&T domains while increasing diversity in the boundary-spanning network could then target to increase the small overlap between the RET and EECS with incentives for focused research at their intersection, given the good representation of the RET and EECS in general (Figure 5). By taking into consideration relative regional strengths in collaborative R&D in both S&T domains, as shown in Figure 3, a component of regional specialization can be readily introduced in policy design.

|  | 2002-2007 | 2005-2010 | 2008-2013 |
|---|---|---|---|
| Authors | 92 | 143 | 237 |
| Linkages | 267 | 323 | 581 |
| Clusters (maximum size) | 19 (21) | 23 (23) | 36 (27) |
| Collaborators per author, average | 5.8 | 4.5 | 4.9 |
| Density | 0.064 | 0.032 | 0.021 |
| $V_t$ | 0.42 | 0.50 | 0.52 |
| $H_t$ | 0.27 | 0.32 | 0.40 |

Table 2. Time-evolution of the boundary-spanning coauthorship network.

To obtain a better understanding of the structure of information in the boundary-spanning network and the broader environment that boundary-spanning research is conducted we identify the most commonly used research terms in communication of research findings, which support information sharing and representation of knowledge at the intersections of scientific fields (Gruber 1995). To this end, we retrieve the keywords associated with these publications of the combined publication records pertaining to of all three S&T domains to which at least one author from the boundary-spanning network of Figure 5 has contributed. Over the period 2010-2015, the five most commonly used keywords are, in order of decreasing frequency, "thin film", "performance", "nanoparticle", "system" and "device". The keywords reflect the level of abstraction of the terms used to define the S&T domains and they suggest significant research activity on the design and optimization of materials and devices/systems in particular. Other very common keywords are "optical property", "polymer", "graphene", "carbon nanotube", and "Raman spectroscopy".

The organizational structure and culture within which researchers are embedded and boundary-spanning research activity is performed affects the nature and direction of the later. Because interactions between individual researchers may be interpreted as interactions between research organizations, it would be convenient to map the boundary-spanning network into a network of organizations, which would enable the study of the role of individual (boundary-spanning) organizations, institutions and funding instruments for R&D in the national innovation system (Edquist 1997). Despite the fact that no giant component emerges in the boundary-spanning network at the level of individual scientists and independent research teams (Figure 5), extended research clusters could exist at the coarser level of organizations. Such investigation remains however beyond the scope of the present study and it will be the topic of a future discussion. Instead, the previous subset of publications was used to perform statistics: we find a relatively high contribution of 46% for GRCs over the period 2010-2015, in accordance with their demonstrated openness (compare with Figure 1b). Relatively high share of European copublication is identified, at 42% over the same period.

Priorities in research collaboration are often defined in terms of funding instead of market opportunities. RDI policies designed to strengthen the cooperation between universities, research organizations, and private firms are used to promote a route to innovation by effectively bridging R&D, demonstration and deployment, and by distributing the risks that often accompany large-scale projects, so it becomes of interest to examine public-private interactions more closely (Etzkowitz 2002). The share of publications in the combined set of records of all three S&T domains to which at least one private organizations has contributed was 5% over the period 2010-2015. We also find the share of HEIs and GRC contributions in such copublications to be 42% (34% over the period 2000-2005) and

15% (5%) of the total number of publications, respectively (compare with Figure 1b). The share of publications over the period 2010-2015 to which additionally an author in the boundary-spanning network has contributed is found to be statistically insignificant. This weak link points at potential imbalances in the way that researchers and research organizations correspond to the needs of private firms in energy-related research knowledge and can have a detrimental effect on ability of firms to absorb external knowledge. After inspection, most firms were found to come from the EECS domain but increased knowledge accumulation and diversity in the boundary-spanning network can encourage new entrepreneurial initiatives, given that appropriate incentives exist to direct the interest of firms towards the research priorities signaled by the boundary-spanning network. We nevertheless recognize that scholarly co-publication is not necessarily a major mechanism for cultivating interactions between the public and private sectors.

# 5. Conclusions

We introduced a combined statistical and network-based approach to study scientific publishing output. We used it to map collaboration patterns in research within the domain of renewable energy technology and its intersections with the domains of nanoscience and nanotechnology with focus on materials, and electrical engineering and computer science in Greece and its broader European and international environment as a case study. Our analysis concerns the sixteen-year period 2000-2015 and includes a description of the S&T domains, demonstrating how it can support evidence-based monitoring of research activity. The three S&T domains characterized and studied in this work are not only of importance for energy technology innovation, but they can also support the development of innova-tive products and processes in other sectors, from transport to tourism to cultural heritage.

Our methods allowed us to establish a (positive) relationship between expenditure on R&D at the national level and the extent and diversity of team-based energy-related research at the intersections of the three S&T domains, which suggests that knowledge creation and diversification is sensitive to broader funding opportunities. Using binary operations on the corresponding coauthorship networks, we found only very few authors to be active at the intersection of all three S&T domains and that increased diversity is associated with a larger number of mostly independent research teams. The fact that the corresponding research network comprises mostly disconnected clusters at all times is indicative of barriers to the advancement of deliberate interdisciplinary research. Funding initiatives that target to create linkages between distinct but complementary research themes can contribute towards tighter integration of R&D efforts that transgress the boundaries of independent research teams. Our findings suggest that such policy initiatives could, for instance, target specifically the interface of the domains of renewable energy technology and electrical engineering and computer science. Additionally, the identification of regional research hubs in interregional and international collaboration which support the two knowledge bases and are in close proximity can inform policies for regional RDI. Care must be taken for the correct interpretation of our case study's findings. Our intention has been to demonstrate our approach and provide examples of how it can be used to identify points of intervention for RDI policy rather than offer policy solutions. Owing to the complexity of RDI systems, interpretation should be attempted only within the provided contextual framework and preferably alongside other complementary, albeit not derived, quantitative descriptions of the RDI system as well as qualitative descrip-tions of it (for example using surveys). Nevertheless, although a detailed examination the transferability of our findings pertaining to Greek research networks to other national RDI systems is beyond the scope of this work, we expect that they will be relevant to systems of similar extend, within which the main contributor to GERD is the public sector, and when there is an interest in incentivizing interdisciplinary research as a means to promote innovation. It is emphasized that coauthorship describes only partially the frequency of interactions between researchers or the nature of their contributions within a scientific discipline. No distinction between negative or positive citation has been attempted here; neither *Web of Science* offers a complete record of scholarly publication.

The static and dynamic descriptions of the international boundary-spanning network and constituent research clusters, as well as our simple concepts and quantitative indicators of knowledge creation and diversification concern RDI policy for energy technology specifically, however our approach can be useful for operationalizing boundary-spanning research and for designing, monitoring, and evaluating interdisciplinary research programs on the basis of empirical evidence more generally. With the particular focus on international collaboration, the implications of our specific findings necessarily extend beyond national borders, while our methods can be applied to other organizational or sectoral, regional or national systems of innovation. A possible direction for future work is the assessment of the degree of integration of research efforts at the coarser level of (boundary-spanning) organizations towards the formation of an extended research cluster and its institutionalization, in order to obtain insights into the relative intensity and pace of knowledge integration at different levels of segregation. Another possibility is to delve deeper into the funding sources and flows at a level of detail finer than the national and this way shed light into the impact of funding incentives on the organization of interdisciplinary research. In any case, even a well-functioning research network can only partially compensate for possible structural deficiencies in the broader innovation system (business, regulatory, and other environments), and therefore it alone is not sufficient to bring technological change or to promote regional development and innovation-driven growth. Further investigation will be necessary to identify such structural deficiencies and associated barriers to knowledge creation and diffusion, and technological learning.

## Acknowledgements

GAT wishes to thank Efthimios Kaxiras at Harvard University Department of Physics, and the members of the Science, Technology, and Public Policy program at the Belfer Center for Science and International Affairs, Harvard Kennedy School for useful discussions and feedback.

## Funding

GAT acknowledges support by a postdoctoral fellowship from the Institute for Applied Computational Science at Harvard John A. Paulson School of Engineering and Applied Sciences.

## References

Aguiléra, A., Lethiais, V., & Rallet, A. (2012). 'Spatial and Non-spatial Proximities in Inter-firm Relations: An Empirical Analysis', *Industry & Innovation*, 19/3: 187–202. DOI: 10.1080/13662716.2012.669609

Aldrich, H., & Herker, D. (1977). 'Boundary Spanning roles and Organization Structure.', *Academy of Management Review*, 2/2: 217–30. DOI: 10.5465/AMR.1977.4409044

Arico, A. S., Bruce, P., Scrosati, B., Tarascon, J.-M., & van Schalkwijk, W. (2005). 'Nanostructured materials for advanced energy conversion and storage devices', *Nat. Mater.*, 4/5: 366–77. DOI: 10.1038/nmat1368

Arora, S. K., Porter, A. L., Youtie, J., & Shapira, P. (2013). 'Capturing new developments in an emerging technology: an updated search strategy for identifying nanotechnology research outputs', *Scientometrics*, 95/1: 351–70. DOI: 10.1007/s11192-012-0903-6

Arundel, A., & Hollanders, H. (2008). 'Innovation scoreboards: indicators and policy use'. *Innovation Policy in Europe: Measurement and Strategy*.

Audretsch, D. B., & Feldman, M. P. (1996). 'R&D spillovers and the geography of innovation and production', *The American Economic Review; Nashville*, 86/3: 630.

Baños, R., Manzano-Agugliaro, F., Montoya, F. G., Gil, C., Alcayde, A., & Gómez, J. (2011). 'Optimization methods applied to renewable and sustainable energy: A review', *Renewable and Sustainable Energy Reviews*, 15/4: 1753–66. DOI: 10.1016/j.rser.2010.12.008

Bernal-Agustín, J. L., & Dufo-López, R. (2009). 'Simulation and optimization of stand-alone hybrid renewable energy systems', *Renewable and Sustainable Energy Reviews*, 13/8: 2111–8. DOI: 10.1016/j.rser.2009.01.010

Bettencourt, L. M. A., & Kaur, J. (2011). 'Evolution and structure of sustainability science', *Proceedings of the National Academy of Sciences*, 108/49: 19540–5. DOI: 10.1073/pnas.1102712108

Boschma, R. (2005). 'Proximity and Innovation: A Critical Assessment', *Regional Studies*, 39/1: 61–74.

Bottasso, C. L., Campagnolo, F., Croce, A., Dilli, S., Gualdoni, F., & Nielsen, M. B. (2014). 'Structural optimization of wind turbine rotor blades by multilevel sectional/multibody/3D-FEM analysis', *Multibody System Dynamics*, 32/1: 87–116. DOI: 10.1007/s11044-013-9394-3

Breschi, S., Lissoni, F., & Malerba, F. (2003). 'Knowledge-relatedness in firm technological diversification', *Research policy*, 32/1: 69–87.

Bresnahan, T. F., & Trajtenberg, M. (1995). 'General purpose technologies "Engines of growth"?', *Journal of Econometrics*, 65/1: 83–108. DOI: 10.1016/0304-4076(94)01598-T

Cooke, P. (2001). 'Regional Innovation Systems, Clusters, and the Knowledge Economy', *Industrial and Corporate Change*, 10/4: 945–74. DOI: 10.1093/icc/10.4.945

Cronin, B., & Sugimoto, C. R. (Eds). (2014). *Beyond bibliometrics: harnessing multidimensional indicators of scholarly impact*. Cambridge, Massachusetts: The MIT Press.

Cummings, J. N., & Kiesler, S. (2005). 'Collaborative Research Across Disciplinary and Organizational Boundaries', *Social Studies of Science*, 35/5: 703–22. DOI: 10.1177/0306312705055535

Edquist, C. (1997). *Systems of Innovation: Technologies, Institutions, and Organizations*. Psychology Press.

Etzkowitz, H. (2002). 'Incubation of incubators: innovation as a triple helix of university-industry-government networks', *Science and Public Policy*, 29/2: 115–28. DOI: 10.3152/147154302781781056

European Commission. (2014). 'A policy framework for climate and energy in the period from 2020 to 2030 [COM(2014) 15]'.

Freeman, L. C. (1977). 'A Set of Measures of Centrality Based on Betweenness', *Sociometry*, 40/1: 35. DOI: 10.2307/3033543

Frenken, K., Oort, F. V., & Verburg, T. (2007). 'Related Variety, Unrelated Variety and Regional Economic Growth', *Regional Studies*, 41/5: 685–97. DOI: 10.1080/00343400601120296

Godin, B. (2006). 'The Linear Model of Innovation: The Historical Construction of an Analytical Framework', *Science, Technology, & Human Values*, 31/6: 639–67. DOI: 10.1177/0162243906291865

Grätzel, M. (2005). 'Solar Energy Conversion by Dye-Sensitized Photovoltaic Cells', *Inorganic Chemistry*, 44/20: 6841–51. DOI: 10.1021/ic0508371

Gruber, T. R. (1995). 'Toward principles for the design of ontologies used for knowledge sharing?', *International Journal of Human-Computer Studies*, 43/5–6: 907–28. DOI: 10.1006/ijhc.1995.1081

Guimerà, R., Uzzi, B., Spiro, J., & Amaral, L. A. N. (2005). 'Team Assembly Mechanisms Determine Collaboration Network Structure and Team Performance', *Science*, 308/5722: 697–702. DOI: 10.1126/science.1106340

Hagberg, A. A., Schult, D. A., & Swart, P. J. (2008). 'Exploring Network Structure, Dynamics, and Function using NetworkX'. Varoquaux G., Vaught T., & Millman J. (eds) *Proceedings of the 7th Python in Science Conference*, pp. 11–5. Pasadena, CA USA.

Hidalgo, C. A., & Hausmann, R. (2009). 'The building blocks of economic complexity', *Proceedings of the National Academy of Sciences*, 106/26: 10570–5. DOI: 10.1073/pnas.0900943106

Hollanders, H., & Es-Sadki, N. (2016). *European Innovation Scoreboard*.

Huang, C., Notten, A., & Rasters, N. (2011). 'Nanoscience and technology publications and patents: a review of social science studies and search strategies', *The Journal of Technology Transfer*, 36/2: 145–72. DOI: 10.1007/s10961-009-9149-8

Huynh, W. U. (2002). 'Hybrid Nanorod-Polymer Solar Cells', *Science*, 295/5564: 2425–7. DOI: 10.1126/science.1069156

International Energy Agency. (2016). *World Energy Outlook*. (Organization for Economic Cooperation and Development, Ed.). Paris: Organization for Economic Co-operation and Development.

Jacobs, J. (2009). 'Interdisciplinarity: A Critical Assessment', *Annual Review of Sociology*, 35: 43–65.

Jovanovic, B., & Rousseau, P. L. (2005). 'General Purpose Technologies'. *Handbook of Economic Growth*, Vol. 1, pp. 1181–224. Elsevier. DOI: 10.1016/S1574-0684(05)01018-X

Kajikawa, Y., Yoshikawa, J., Takeda, Y., & Matsushima, K. (2008). 'Tracking emerging technologies in energy research: Toward a roadmap for sustainable energy', *Technological Forecasting and Social Change*, 75/6: 771–82. DOI: 10.1016/j.techfore.2007.05.005

Luukkonen, T., Persson, O., & Sivertsen, G. (1992). 'Understanding Patterns of International Scientific Collaboration', *Science, Technology, & Human Values*, 17/1: 101–26. DOI: 10.1177/016224399201700106

Manning, C. D., Raghavan, P., & Schütze, H. (2008). *Introduction to Information Retrieval.*, 1 edition. New York: Cambridge University Press.

Manzano-Agugliaro, F., Alcayde, A., Montoya, F. G., Zapata-Sierra, A., & Gil, C. (2013). 'Scientific production of renewable energies worldwide: An overview', *Renewable and Sustainable Energy Reviews*, 18: 134–43. DOI: 10.1016/j.rser.2012.10.020

Milojević, S. (2014). 'Principles of scientific research team formation and evolution', *Proceedings of the National Academy of Sciences*, 111/11: 3984–9. DOI: 10.1073/pnas.1309723111

Mogoutov, A., & Kahane, B. (2007). 'Data search strategy for science and technology emergence: A scalable and evolutionary query for nanotechnology tracking', *Research Policy*, 36/6: 893–903. DOI: 10.1016/j.respol.2007.02.005

Newman, M. (2010). *Networks: An Introduction*. Oxford, New York: Oxford University Press.

Newman, M. E. J. (2004). 'Coauthorship networks and patterns of scientific collaboration', *Proceedings of the National Academy of Sciences*, 101/suppl 1: 5200–5. DOI: 10.1073/pnas.0307545100

Nonaka, I. (1994). 'A Dynamic Theory of Organizational Knowledge Creation', *Organization Science*, 5/1: 14–37. DOI: 10.1287/orsc.5.1.14

Porter, A. L., & Youtie, J. (2009). 'How interdisciplinary is nanotechnology?', *Journal of Nanoparticle Research*, 11/5: 1023–41. DOI: 10.1007/s11051-009-9607-0

Rafols, I., & Meyer, M. (2010). 'Diversity and network coherence as indicators of interdisciplinarity: case studies in bionanoscience', *Scientometrics*, 82/2: 263–87. DOI: 10.1007/s11192-009-0041-y

Romer, P. M. (1986). 'Increasing Returns and Long-Run Growth', *Journal of Political Economy*, 94/5: 1002–37. DOI: 10.1086/261420

——. (1990). 'Endogenous Technological Change', *Journal of Political Economy*, 98/5, Part 2: S71–102. DOI: 10.1086/261725

Sharif, N. (2006). 'Emergence and development of the National Innovation Systems concept', *Research Policy*, 35/5: 745–66.

de Solla Price, D. J. (1965). 'Networks of Scientific Papers', *Science*, 149: 510–5. DOI: 10.1126/science.149.3683.510

UNESCO (Ed.). (2015). *UNESCO Science Report: towards 2030*. Paris: United Nations Educational Scientific and Cultural.

Wang, C.-Y. (2004). 'Fundamental Models for Fuel Cell Engineering', *Chemical Reviews*, 104/10: 4727–66. DOI: 10.1021/cr020718s

Weitzman, M. L. (1998). 'Recombinant Growth', *The Quarterly Journal of Economics*, 113/2: 331–60. DOI: 10.1162/003355398555595

Yu, W., An, D., Griffith, D., Yang, Q., & Xu, G. (2015). 'Towards Statistical Modeling and Machine Learning Based Energy Usage Forecasting in Smart Grid', *SIGAPP Appl. Comput. Rev.*, 15/1: 6–16. DOI: 10.1145/2753060.2753061

# Interdisciplinary collaboration in research networks: Empirical analysis of energy-related research in Greece

**Georgios A. Tritsaris[1,\*], and Afreen Siddiqi[2,3]**

[1]*Harvard John A. Paulson School of Engineering and Applied Sciences, Harvard University, Cambridge, MA 02138, USA*

[2]*Belfer Center for Science and International Affairs, John F. Kennedy School of Government, Harvard University, Cambridge, MA 02138, USA*

[3]*Institute for Data, Systems, and Society, Massachusetts Institute of Technology, Cambridge, MA 02139, USA*

*Corresponding author. georgios@tritsaris.eu

## Data and methods

This study covers the sixteen-year period 2000-2015 and relies on publication and citation data available through the *Web of Science* (*WoS*) platform and its core selection of citation indexes and databases. Publication and citation indexes covering the social sciences and humanities were excluded to concentrate attention on the technical aspects of the three S&T domains considered here. Only journal articles and proceedings papers with at least one author in affiliation with a Greek organization were used. For each publication the record of citing publications for the same sixteen-year period was also retrieved using the *Web of Knowledge Web Services* (journal articles and proceedings papers only). All data was stored and managed using the *MySQL* relational database system (release 5.6). Hereafter, we will refer to the complete set of records simply as the "collection" ("EL" label in figures). For the purposes of this study, we developed software for data collection, cleaning, analysis and visualization, which supports complex queries such as *"retrieve all publications under the subject category of Computer Science, which are the product of an international collaboration, and which have been cited at least one time the year of publication or the following year."*

   Publications within the RET and NNM domains were identified in the collection using simple keyword-based queries with equal weight on publication titles, abstracts and associated

keywords (Manning et al. 2008). The text was normalized before any attempt in retrieval: after breaking up all text into tokens (or words) each noun in plural form was replaced by its singular form as a means of token normalization. Singularization of nouns was preferred to stemming in order to better exploit the precision of scientific writing for more precise retrieval (fraction of relevant publications among the retrieved publications). These operations were performed using the Python open-source library *Natural Language Toolkit* (NLTK, version 3.2 (Bird et al. 2009)). For each of the two S&T domains a set of handpicked terms was used to retrieve an initial set of publications and the keywords associated with them. Keywords with total number of occurrences less than that of the highest-count keyword associated with a cumulative distribution function of 0.05 were discarded. Separate sets of publications were retrieved using the remaining keywords, one at a time. A keyword was considered for inclusion in the initial set of handpicked terms if at least one of the latter appeared among the most frequent keywords associated with the corresponding set of publications. Some keywords were eliminated after manual inspection either as irrelevant or as of inappropriate level of abstraction on the basis of our domain expertise: for example keywords such as "system" or "water" were not considered. The above procedure serves therefore as a form of query expansion which is used to improve recall (fraction of relevant publications that have been retrieved over total relevant publications) by refining an initial set of handpicked terms used to describe a S&T domain, while new and emerging areas of inquiry are probed by careful selection of the initial set of keywords.

The collection of the sixteen-year period between 2000 and 2015 contains in total 154339 publications. Publications with a number of authors greater than 100 were not considered. This constraint improved the robustness of our statistical analysis by excluding work of large collaborative projects such as those in high-energy physics, which generally remain disconnected from the S&T domains of interest. After data cleaning and integrity checks the total number of publications are 152124 with 124010 journal articles and 28114 conference proceedings.

Bird, S., Klein, E., & Loper, E. (2009). *Natural Language Processing with Python: Analyzing Text with the Natural Language Toolkit.* O'Reilly Media, Inc.

Manning, C. D., Raghavan, P., & Schütze, H. (2008). *Introduction to Information Retrieval.*, 1 edition. New York: Cambridge University Press.

The query terms used for defining the RET and NNM domains are provided below.

## Query terms for renewable energy technology (RET)

The query terms used describe renewable energy sources, related technologies for energy conversion ("solar panel", "wind turbine"), and fuels ("hydrogen production", "biofuel"). Energy sources with lesser significance in the national energy supply, such as marine and geothermal, were not considered. Most common keywords co-occurring with the elected terms include the general terms "energy", "system", and "performance" (these were discarded).

bio oil, biodiesel, bioenergy, biofuel, biomass energy, biomass fuel, biomass power, clean energy, clean power, fuel cell, green energy, green power, hydrogen energy, hydrogen fuel, hydrogen production, hydrogen storage, methanol fuel, photoelectrochemical cell, photovoltaic solar energy, renewable energy, renewable energy source, renewable power, solar cell, solar collector, solar concentrator, solar electricity, solar fuel, solar panel, solar power, solar thermal, wind electricity, wind energy, wind power, wind turbine.

**Query terms for nanoscience and nanotechnology with focus on materials (NNM)**
The query terms used to describe publications in the NNM domain describe common objects of study ("quantum dot", "micelle", "nanodevice") and their morphology ("monolayer", "mesopore"). We excluded terms specific to synthesis and characterization techniques, which are nevertheless commonly discussed in studies of materials. Most frequently co-occurring keywords include "surface" and "spectroscopy" (these general terms were discarded).

atomic structure, electronic structure, graphene, mesopore, mesoporous, micelle, molecular structure, monolayer, nanocluster, nanocomposite, nanocrystal, nanocrystalline, nanodevice, nanoelectronic, nanofiber, nanomaterial, nanoparticle, nanophotonic, nanopore, nanoporous, nanorod, nanoscience, nanostructure, nanostructured, nanosystem, nanotechnology, nanotube, nanowire, quantum dot, thin film.

**Subject categories for electrical engineering and computer science (EECS)**
Electrical and Electronic Engineering, and all subcategories within Computer Science.

## Data on research inputs

|        | 2000   | 2001   | 2002    | 2003    | 2004    | 2005    | 2006    | 2007    |
|--------|--------|--------|---------|---------|---------|---------|---------|---------|
| **Greece** | :      | 852    | :       | 978     | 1021    | 1154    | 1223    | 1342    |
| **EU-28**  | 171197 | 179346 | 186807  | 188827  | 194341  | 202129  | 216330  | 229582  |
|        | 2008   | 2009   | 2010    | 2011    | 2012    | 2013    | 2014    | 2015    |
| **Greece** | 1602   | 1486   | 1353    | 1391    | 1338    | 1466    | 1489    | 1684    |
| **EU-28**  | 239942 | 237421 | 246995  | 259892  | 270322  | 274500  | 286121  | 298811  |

Gross domestic expenditure on R&D (GERD, million euro), source: Eurostat

|        | 2000    | 2001    | 2002    | 2003    | 2004    | 2005    | 2006    | 2007    |
|--------|---------|---------|---------|---------|---------|---------|---------|---------|
| **Greece** | :       | 30226   | :       | 31849   | :       | 33603   | 35140   | 35531   |
| **EU-28**  | :       | :       | 2087106 | 2106121 | 2146198 | 2201518 | 2293678 | 2370179 |
|        | 2008    | 2009    | 2010    | 2011    | 2012    | 2013    | 2014    | 2015    |
| **Greece** | :       | :       | :       | 36913   | 37361   | 42188   | 43316   | 50512   |
| **EU-28**  | 2463973 | 2488453 | 2541908 | 2612979 | 2670841 | 2712854 | 2774612 | 2848841 |

Full-time equivalent (FTE), source: Eurostat

|        | 2000    | 2001    | 2002    | 2003    | 2004    | 2005    | 2006    | 2007    |
|--------|---------|---------|---------|---------|---------|---------|---------|---------|
| **Greece** | :       | 55626   | :       | 56708   | :       | 61454   | :       | :       |
| **EU-28**  | :       | :       | :       | 3029939 | :       | 3198490 | :       | 3451996 |

| | 2008 | 2009 | 2010 | 2011 | 2012 | 2013 | 2014 | 2015 |
|---|---|---|---|---|---|---|---|---|
| **Greece** | : | : | : | 70229 | : | 82684 | : | : |
| **EU-28** | : | 3662671 | 3793294 | 3967636 | : | 4154568 | : | : |

Head count (HC), source: Eurostat

| Administrative region (NUTS2) | GERD |
|---|---|
| **Attica** | 820.27 |
| **Central Greece** | 35.28 |
| **Central Macedonia** | 183.30 |
| **Crete** | 120.68 |
| **Eastern Macedonia and Thrace** | 43.21 |
| **Epirus** | 39.78 |
| **Ionian Islands** | 8.17 |
| **North Aegean** | 21.40 |
| **Peloponnese** | 30.82 |
| **South Aegean** | 14.98 |
| **Thessaly** | 50.27 |
| **Western Greece** | 79.72 |
| **Western Macedonia** | 17.80 |

Regional GERD 2013 (million euro), source: National Documentation Center

 "Research & Development Expenditure and Personnel in Greece in 2013", National Documentation Center, 2015
"Research, Development & Innovation in Greek regions", National Documentation Center, 2015

# Types of organizations

The following categorization is used for organizations:

**1. Higher Education Institutions – Universities**
Public universities and technical universities.
**2. Higher Education Institutions – Technological Education Institutes**
Technological Education Institutes and the Higher School of Pedagogical and Technological Education (ASPAITE).
**3. Higher Education Institutions – Other**
The National School of Public Health, military academies, ecclesiastical schools, and others.
**4. Public Research Centers – supervised by the General Secretariat of Research and Technology (GSRT)**
The National Observatory of Athens, the Foundation for Research and Technology – Hellas (FORTH), the National Center for Scientific Research " DEMOKRITOS", and others.

**5. Public Research Centers – Other**
The Academy of Athens, the Hellenic Agricultural Organization DEMETRA, the Benaki
Phytopathological Institute, Computer Technology Institute and Press "Diophantus", and others.

**6. Other Public Institutions**
Organizations and services of public administration.

**7. Health Institutions – Public**
Public hospitals, clinics and others, including organizations supervised by the Ministry of
National Defense.

**8. Health Institutions – Private**
Private hospitals, clinics, diagnostic centers, research centers, and others.

**9. Private sector – excluding nonprofit organizations**
Organizations and institutions whose primary activity is the market production of goods or
services.

**10. Private sector - Nonprofit organizations**
Professional and learned societies, relief or aid agencies, unions and others.

HEI sector comprises (1), (2), and (3). GRC sector comprises (4), (5), and (6).

# Additional findings

## National publishing output
Although this study focuses on trends in national research output across S&T domains, it is still of
interest to compare national output with this of the other EU-28 member states. National
publishing output has been between 47% and 73% of the EU-28 average (Figure 1S), however it is
raised to around 400% after taking into account national gross domestic expenditure on R&D
(GERD). The gap narrows if normalization is performed with respect to full-time equivalent (FTE)
or head count (HC). For the FTE and HC only partial data is available, highlighting the need for
improving the data collection process, although the same trend as GERD can be assumed. We did
not perform detailed comparisons for the three S&T domains across national borders as this
would be beyond the scope of this work.

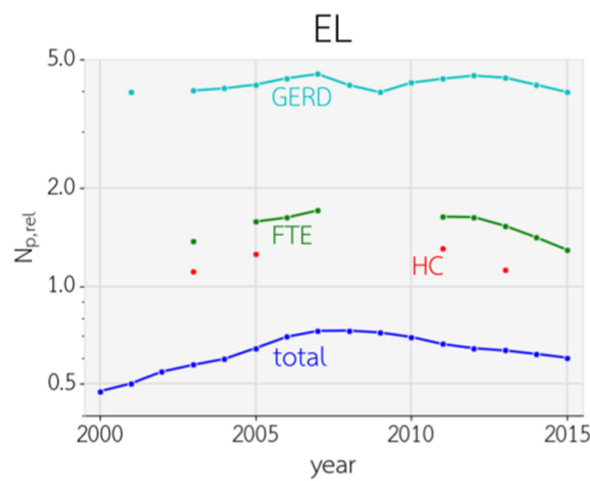

Figure 1S. National publishing output relatively to EU-28 average with respect to total number of publications (blue), number of publications per gross domestic expenditure on R&D (cyan), full-time equivalent (green), and head count (red).

## Research impact of S&T domains

Alongside scholarly publication, citations are often treated as measures to research output, although not necessarily as an indicator of novelty and impact at the forefront of a research field. Figure 2S shows the citation impact (CI) per type of collaboration for each S&T domain. We define the citation impact for a given year as the ratio of the number of citations recorded that year to the total number of publications of the preceding two years. This definition renders CI a time-sensitive indicator. To account for the different citation practices across S&T disciplines and time periods, CI is normalized by dividing with the domain average over the same span of years and for each *WoS* subject category ($CI_{rel}$). Here, self-citations are included. The two types of collaboration we considered involve a) at least one EU-28 member state, regardless of any other contributions ("European" in Figure 2S-b), or b) at least one country outside the EU-28, regardless of EU-28's participation ("International"). Published research without the contribution of any other country is also provided for comparison ("National"). The most obvious trend across all domains is the higher impact of collaborative research ($CI_{rel} > 1$). The apparent convergence of the CI indicators within the EECS domain can be attributed to the continuing expansion and internationalization of the domain, which would bring the domain average closer to the "European" and "International" trends and therefore decrease the respective values for $CI_{rel}$ (see also Figures 1a and 1c).

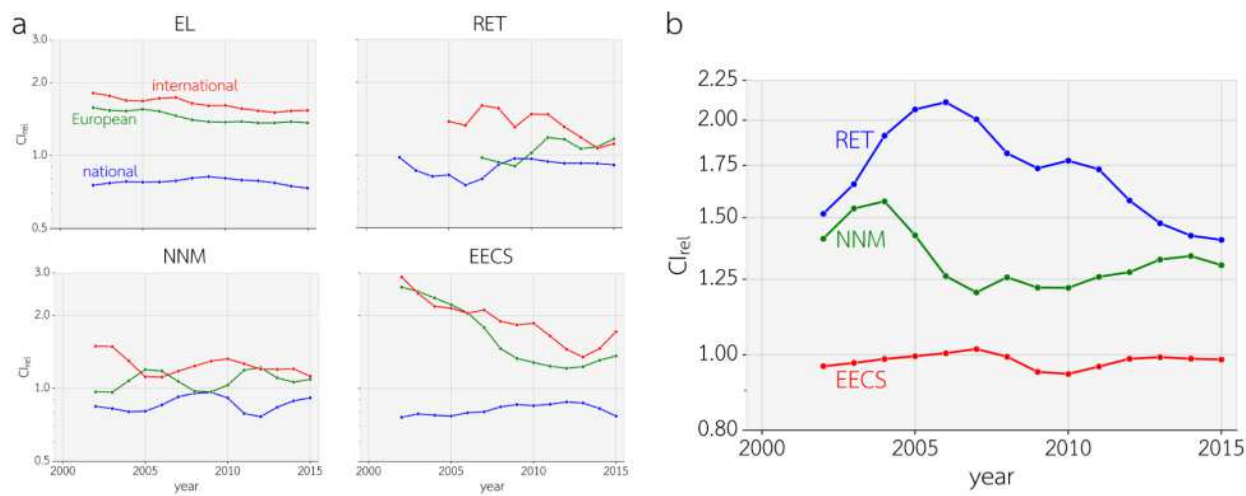

Figure 2S. a) Citation impact by type of collaboration in scientific publication relatively to domain-specific average. b) Citation impact by science and technology domain relatively to national average.

The citation impact of each of the three S&T domains with respect to the collection average was also calculated (Figure 2S-b). The impact of the RET and NNM domains is found to be above average. The impact of the EECS domain follows closely the national average, which can be attributed to the relatively large number of publications assigned to the domain (see also Figure 1a).To understand the impact of self-citations on the calculated indicators, we calculated a rate for self-citation for a given year for each S&T domain as the number of publications citing another the year of its publication or the following year over the total number of publications that year. The rate has been between 20% and 25% for most of the period 2000-2015 for the RET and NNM domains, while it has been steadily decreasing from 40% for the EECS domain towards 20%.